\documentstyle[sprocl,epsfig]{article}

\def\be{\begin{equation}}

\def\ee{\end{equation}}

\def\bea{\begin{eqnarray}}

\def\eea{\end{eqnarray}}

\begin{document}

\title{WAVELET PERSPECTIVE OF A DISORIENTED \\ CHIRAL CONDENSATE 
\footnote{Talk 
presented
by Ina Sarcevic at DPF '96 and by Zheng Huang at PANIC '96}}

\author{Zheng Huang$^\dagger$,
Ina Sarcevic$^\dagger$, 
Robert Thews$^\dagger$ and
Xin-Nian Wang$^\ddagger$}

\address{{$^\dagger$Department of Physics,
\baselineskip=12pt University of Arizona, Tucson, AZ 85721}\\
{$^\ddagger$Nuclear Science Division,
LBNL, Berkeley, CA 94720}}




\vskip -0.1in
\maketitle\abstracts{
The possible chiral phase transition in high energy heavy-ion
collisions may lead to
 the formation of a disoriented chiral condensate (DCC).
However, the existence of  many uncorrelated small domains in
the rather large interaction volume and the huge combinatorial 
background make the experimental search for the DCC signal a 
rather difficult task. 
We present a novel method for  studying  the
formation of a DCC in
high energy heavy-ion collisions utilizing
a discrete wavelet transformation. 
We find the wavelet power spectrum for
a DCC to exhibit a strong dependence on the scale while
an equilibrium system and  the standard dynamical
models such as HIJING have a flat spectrum.}

\vskip -0.1in

In heavy-ion collisions 
the rapid expansion of the collision debris 
in the longitudinal  direction may lead to supercooling  
of the interaction region, and as a result, domains of the
``unconventionally'' oriented vacuum configurations allowed by the   
chiral symmetry may be formed.\cite{dcc} 
Detection of this interesting 
phenomenon, the so-called Disoriented Chiral Condensate (DCC) \cite{dcc}, 
would provide valuable information on the vacuum structure of the strong 
interaction and the nature of the chiral phase transition.  
Preliminary theoretical investigations on non-equilibrium dynamics 
using the classical linear $\sigma$-model have found some evidence 
for the growth of long wavelength pion modes, which may indeed lead 
to  domain structure or  cluster formation.\cite{rw,hw} 
If there are many uncorrelated small domains, the integrated probability 
distribution of the neutral pion fraction $f$ emitted from 
a disoriented region, predicted to be $P(f)=1/2\sqrt f$, \cite{dcc} 
would become  Gaussian.
This 
makes the experimental search for the DCC signal a rather difficult task.  
In order to disentangle the DCC domain structure in high energy heavy-ion
collisions, 
we propose a new 
method which emphasizes not only the behavior of the probability 
distribution in the full phase space region but also its
 fluctuation in rapidity $\eta$ or azimuthal angle $\phi$. 
In other words, one needs to study the ``local'' properties of the 
distribution in phase space  if  the DCC clusters are
``localized'' objects in phase space.  

A novel method 
is 
a multi-resolution analysis performed by a discrete wavelet
transformation (DWT) 
\cite{wavelet} 
which has been found effective in systematically
detecting structures of various scales in turbulence,
and multiparticle productions.\cite{farge,carr} 
For any one-dimensional sample of a point distribution
$f(x)$, such as  the rapidity distribution of the neutral pion fraction
in the interval $[0,1]$ with resolution $\Delta x$, can be represented
as a histogram of $2^J$ bins where $J={\rm mod}(|{\rm ln}
\Delta x|/{\rm ln}2)+1$ 
\begin{equation}
f(x)\equiv  f^{(J)}(x)=\sum_{k=0}^{2^J-1}f_{Jk}\phi_{Jk}(x), \label{1}
\end{equation}
where $\phi_{Jk}(x)$ are the piecewise constant functions called the
mother functions. The 
histogram function at the next
finer resolution is the average of adjacent bins at the finest scale.
Evidently some detail is lost compared to (\ref{1}). The difference
between (\ref{1}) and $f^{(J-1)}$ can be fully expressed in terms
of the diffence functions $\psi _{J-1,k}(x)$ which are stairwell functions
called father functions. The difference amplitudes or   
one defines the DWT coefficients 
\begin{eqnarray}
\tilde{f}_{jk}=2^j\int f(x)\psi_{jk}(x)dx,\label{6b}
\end{eqnarray}
where the basis functions $\psi_{jk}(x)$ have a unique property: they
are related to each other by scaling and translation:
$\psi_{jk}(x)=\psi (2^jx-k)$.
The index $j$ denotes the resolution scale and $k$
the position at each scale. The idea of multiresolution analysis is to find
representations of the sample function $f(x)$ at various scales. 
Unlike the Fourier transformation which
requires the information in whole physical space, the wavelet SSD requires
only the local information in space. 
Roughly speaking, the multiresolution
analysis is a journey down (or up) a hierarchy of scales and to view
the ``world'' in different ``eyes'' with magnifying or reducing glasses.

The central idea of using the wavelet analysis is to attempt an unbiased
separation of the DCC physics from the various ``backgrounds'' which cannot
be otherwise easily eliminated. The usual correlation method is not so
effective here because the different scale correlations are mixed together
in the same phase space. The discrete wavelet transformation (DWT), on the
other hand, has an extra dimension -- the scale
which can be used to separate these correlations.
The main source of the ``backgrounds'' is the
fluctuations at small correlation scale due to pion
radiation from the uncorrelated regions in space-time, which 
can coexist with a DCC domain. The background can also come from the
partial thermalization of the system leading to thermal correlations.  
\begin{figure}
\begin{center}
\vskip -0.5in
\epsfig{file=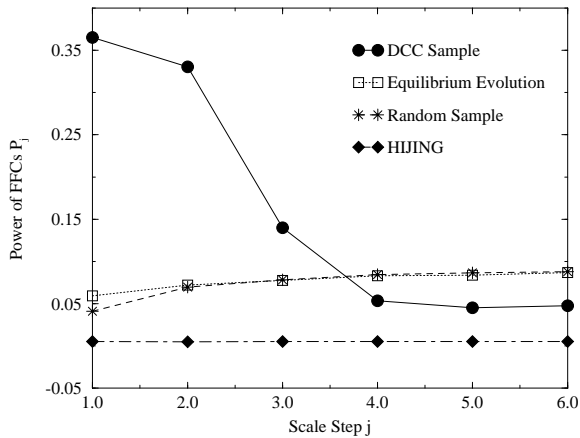,height=2.5in}%
\vskip -0.2in
\caption{The power spectra of the wavelet coefficients for different dynamical
scenarios.}
\end{center}
\end{figure}
We have calculated the wavelet power spectrum for rapidity distribution
of  the neutral pion fraction
simulated using the classical linear $\sigma$
model.\cite{hw}  One defines the power spectrum
with respect to the wavelet basis at each $j$ as
\begin{equation}
P_j=\frac{1}{2^j}\sum_{k=0}^{2^j-1}|\tilde{f}_{jk}|^2\; .
\end{equation}
$P_j$ is then the power of fluctuations on length scale $L/2^j$.
It is
also the mean square of the probability distributions of the 
wavelet coefficients defined in (\ref{6b}). 
The result is 
shown in Fig.1. For the random samples and
the equilibrium evolution of the classical field where there is no DCC
production mechanism, the wavelet power is flat indicating there
is no sensitivity on the change of the scale. For the DCC samples, the
 power starts to build up as $j$ goes down below $j_d\sim 4$ and rises rather
quickly, departing distinctively from the equilibrium and the random
cases. This suggests the existence of the DCC clustering with a typical
size of $\Delta\eta_d=2\eta_{\rm max}/2^{j_d}\simeq 0.8$ units in rapidity.
Also plotted is the power spectrum from HIJING Monte Carlo data which
also features a flat distribution.



\end{document}